\documentstyle[sprocl]{article}

\def\Journal#1#2#3#4{{#1} {\bf #2}, #3 (#4)}

\def\PLB{{\em Phys.Lett.}B}
\def\PRL{\em Phys.Rev.Lett.}
\def\PRD{{\em Phys.Rev.}D}

\def\SJNP{\em Sov.J.Nucl.Phys.}
\def\EPJC{{\em EPJ} C}
\def\JETP{\em JETP}
\def\JETPL{\em JETP Lett.}

\def\ibid{\em ibid}

\def\GC{\em Grav$\&$Cosm}

\def\IV{\em Izv.Vuzov.Fizika}
\def\AJ{\em Ap.J}

\begin{document}

% hep-ph/9902447

\title{Electromagnetic Neutrino Properties and Neutrino Oscillations in
Electromagnetic Fields}

\author{ A.M. EGOROV, A.E. LOBANOV,
A.I. STUDENIKIN\footnote{\normalsize E-mail: studenik@srdlan.npi.msu.su}}

\address{Moscow State University,Department of
Theoretical Physics,119899Moscow,Russia}


\maketitle
%\abstracts{}

%\section{}
%\subsection{}

 {\it Introduction}. The problem of neutrino from the time when
this particle was theoretically recognized (Pauli, 1930; Fermi,
1932-1934) as an important element of the theory of weak interaction,
still is closely connected with the fundamentals of particle physics.
That is why neutrino properties are especially interesting.
However, inspite of great efforts, there is a set of open basic
questions about neutrinos (nonzero mass; number of flavours; mixing
between the neutrino families; behavior under particle-antiparticle
conjugation; electromagnetic properties; the role in astrophysics
etc.). For sure the answer to any of these questions will
substantially stimulate progress in understanding of laws of micro
and macro worlds.

 Electromagnetic moments are among the most important properties of
a particle. Since neutrinos are neutral particles and there is no coupling
to the photon in the lagrangian, electromagnetic moments arise entirely
from vacuum polarization loop effects. For neutrino as a spin-${1\over 2}$
particle the matrix element of electromagnetic current contains four
independent form factors~\cite{0}:
   \begin{equation}
   \begin{array}{c}
   < \nu (p')|J^{em}_{\mu}|\nu (p)>
   ={\vec u}_f \big[ f_{Q}(q^2)\gamma_{\mu}
   +f_{A}(q^2)(q^{2}\gamma_{\mu}-q_{\mu}{\not q})\gamma_5 \\
   +f_{M}(q^2)\sigma_{\mu \nu}q^{\nu}
   +f_{E}(q^2)\sigma_{\mu \nu}q^{\nu}\gamma_5  \big]u_i,~q=p_f - p_i,
   \end{array}
   \end{equation}
where $f_Q$, $f_A$, $f_M$ and $f_E$ are the charge, axial charge,
magnetic and electric dipole moments, respectively.
For Dirac neutrinos the assumption of CP
invariance combined with the hermicity of the electromagnetic
current $J^{em}_{\mu}$
implies that one of two possible static moments, $f_M$ and $f_E$,
vanishes: $f_E=0$.
For neutral Majorana neutrinos from a general assumption of CPT
invariance it follows that both magnetic and electric dipole moments
vanish. However, in the case of the off-diagonal electromagnetic vertex
the neutrino transition magnetic moment is not, in general, zero.

 Thus, one can conclude that the neutrino magnetic (transition) moment
is the most important among the form factors. It should be mentioned
here, that the discussion on the third non trivial electromagnetic
characteristic of neutrino, $f_{A}$, can be found
in \cite{1}.
The magnetic moment arises from the operator $\sigma_{\mu \nu}q^{\nu}$
and since $ {\vec \Psi}'\sigma_{\mu \nu}\Psi
={\vec \Psi}'_{L}\sigma_{\mu \nu}\Psi_{R}
+{\vec \Psi}'_{R}\sigma_{\mu \nu}\Psi_{L}$,
it follows that a chirality change can appear when there are both left
and right-handed particle states. In the standard model due to the
absence of the right-handed charged currents a massless Dirac neutrino
cannot have a magnetic moment.
In the standard model supplied with $SU(2)$-singlet
right-handed neutrino $\nu_{R}$ the one-loop radiative correction
generates nonvanishing magnetic moment, proportional to
$m_{\nu}$~\cite{2,2.1}:  \begin{equation} \begin{array}{c} \mu_{\nu}
   ={3eG_F \over 8\sqrt{2}\pi^2}m_{\nu}
   =3\cdot 10^{-19}({m_{\nu} \over 1 eV})\mu_{0},
   ~~\mu_{0}={e \over 2m_e}.
   \end{array}\label{_} \end{equation}
 There are plenty of
models \cite{3} which predict much larger magnetic moments for
neutrinos. In some of these models \cite{4} the neutrino magnetic
moment can be lifted up to the range $\mu \sim 10^{-10} \div 10^{-12}
\mu_{0}$ that could be of practical interest for processes in the
vecinity of the Sun, neutron star and supernova.

 Experimental upper
bounds on the neutrino magnetic moments are\cite{5}:
$\mu_{\nu_e}\leq 1.8\cdot 10^{-10}\mu_{0}, \mu_{\nu_\mu}\leq 7.4\cdot
10^{-10}\mu_{0}, \mu_{\nu_\tau}\leq 5.4\cdot 10^{-7}\mu_{0}$.
One could also obtain more stringent constraints from astrophysical
arguments like neutrino emission from red giant stars~\cite{6}
or from supernovae and neutron stars~\cite{9_,2PFP,10}.

 {\it Neutrino magnetic moment in electromagnetic field.} It is
non-trivial result shown for the first time in \cite{11_}, that the
presence of medium and external magnetic field change electromagnetic
properties of neutrino. Thus the massless neutrino in the standard model
due to
the weak interaction with particles of the medium can get an
effective magnetic moment. In an analogy with the case of charged
lepton magnetic moment (see, e.g.,~\cite{12}) the non
vanishing intrinsic neutrino magnetic moment also depends on the
strength of external electromagnetic field and the energy of the
neutrino.  The most systematic way in investigation of the external
field and energy dependence of neutrino magnetic moment is based on
consideration of the neutrino mass operator ${\hat M}(x',x)$
in electromagnetic field
that accounts for the field radiative corrections to the neutrino
motion and defines the righthand side of the Schwinger equation
   \begin{equation}
   \begin{array}{c} (i{\not p}-m_{\nu})\Psi_{\nu}(x) =\int {\hat
   M}(x',x,F)\Psi_{\nu}(x'){\rm d}x'.  \end{array}\label{D-SH}
   \end{equation}
We consider the lowest-order standard model contribution (i.e. the
contribution of the virtual loop process $\nu\rightarrow e+W\rightarrow\nu$)
to the Dirac neutrino mass
operator in external crossed electromagnetic field
(${\vec E}\perp {\vec B}$, $E=B$). The details of
calculations of a lepton magnetic moment in the external magnetic
field can be found in \cite{13}.  Finally, for the neutrino magnetic
moment in the external field we find
   \begin{equation}
   \begin{array}{c}
   {\mu^{W}_{\nu}(\chi) \over \mu_{0}} ={g^2 \over 2^5\pi^2}{m_{\nu}
   \over m_{e}} \int\limits^{\infty}_{0} {{\rm d}u \over
   (1+u)^{8/3}}{u^{1/3} \over \chi^{2/3}} [{2(2+u) \over u}
   -{\eta \over \lambda} +{(2u+1) \over \lambda}] \Upsilon (z),
   \eta={m^2_\nu \over m^2_e}, \\ \lambda={m^2_W \over m^2_e},z=({u
   \over \chi})^{2/3}({1+u \over u})^{1/3} (1+{\lambda \over u}-{\eta
   \over 1+u}),\Upsilon (z)=\int\limits^{\infty}_{0}\sin(zx+{x^3
   \over 3}){\rm d}x,
   \end{array}\label{9}
   \end{equation}
where $g$ is the coupling constant of the standard model,
$\chi=[-(eF^{\mu\nu}p_\nu)^2]^{1/2}\cdot \\ \cdot m^{-3}_{e}$ is the field
dynamical parameter, $p_{\nu}$ is the neutrino momentum. The received
expression for $\mu^{W}_{\nu}(\chi)$ exactly accounts for dependence
on the field parameter $\chi$ and also on masses $m_{\nu}$, $m_{e}$,
and $m_{W}$.  If in (\ref{9}) one neglect the terms $\sim{m^2_e \over
m^2_W}$, ${m^2_\nu \over m^2_W}$, then the result of~\cite{ZH} is
achieved.  It is possible to extract the vacuum field and energy
independent part of $\mu^W_\nu$:
   \begin{equation} \begin{array}{c}
   \mu^{W}_{\nu}(0)
   ={g^{2}e \over (8\pi)^2}{m_{\nu} \over m^{2}_{e}}F(\lambda,\eta),\\
   F(\lambda,\eta)={1 \over \eta}\Big\{ 2+{\eta \over \lambda}
   +{1 \over \lambda} +[{1 \over \eta}({1 \over 2\lambda}-\lambda
   +{1 \over 2})-{1 \over 2\lambda}+{3 \over 2}]\ln{\lambda} \\
   +[\eta({3 \over 2}+{1 \over 2\lambda})+{1 \over 2}-{1 \over \lambda}
   -{5 \over 2}\lambda+{1 \over \eta}(\lambda^2-{3 \over 2}\lambda
   +{1 \over 2\lambda})]I(\lambda,\eta) \Big\}, \\
   I(\lambda,\eta)=-{1 \over \sqrt{\Delta}}
   \ln{\epsilon-\sqrt{\Delta} \over \epsilon +\sqrt{\Delta}},
   \Delta=\epsilon^2-4\eta,\epsilon=1+\lambda-\eta.
   \end{array}\label{10}
   \end{equation}
In the case of small $m_{\nu}$,
$\eta\ll1$,
the function $F$ is $F(\lambda,0)={3\lambda-1 \over (\lambda-1)^2}
+{\lambda \over (\lambda-1)^3}\ln{\lambda}$. In
the limit $\lambda\gg1$ we get
$F(\lambda,0)\approx 3/\lambda$, that together
with (\ref{10}) reproduce the result of
eq.(\ref{_}). In analogious way we have also considered~\cite{ES} the
other possible contributions to the neutrino magnetic moment, that
are predicted within extensions of the standard model.

 {\it Neutrino conversion in electromagnetic field.}
Consider a system, $\nu=(\nu_{R},\nu_{L})$,
of two neutrino chiral components in
a magnetic field ${\vec B}$. The evolution of $\nu$ can be written
as a Schr\"odinger
like equation ( see~\cite{9_6} and also~\cite{7})
   \begin{equation}
   \begin{array}{c}
   i{\partial \over \partial t}\nu =H\nu,~~H=({\vec \sigma}{\vec n})
   ({\Delta m^2A \over 4E} -{V \over 2})
   -\mu{\vec \sigma}\Big({\vec B}-{\vec n}({\vec B}{\vec n})\Big),
   \end{array}\label{1}
   \end{equation}
where ${\vec n}$ is the unit vector in the direction of neutrino speed
${\vec u}$, ${\vec \sigma}$ is
the Pauli matrixes, $V$
is difference of neutrino effective potentials in matter, A is
a function of the neutrino vacuum mixing angle $\theta$ which is
determined by the considered type of neutrino conversion
process (for example,
$A={1 \over 2}(\cos{\theta}-1)$ for $\nu_{eL}\leftrightarrow
\nu_{eR}$).
In the Hamiltonian (\ref{1}) the term that is
proportional to the unit matrix is omitted.
Equation (\ref{1}) can be received in the frame of generalized
standard models of electroweak interactions. However, it can be
derived from some general arguments within quaziclassical approach.
It is well known that in the classical approximation the evolution of
particle spin can be determined by the Bargmann-Michel-Telegdi
equation \cite{8}.  On the basis the BMT equation one can
get~\cite{8.1} the following equation for the spin vector $S^{\mu}$
of the neutral particle moving with constant speed, $u_{\mu}=const$,
in electromagnetic field $F_{\mu\nu}$:  \begin{equation}
\begin{array}{c} {dS^{\mu} \over d\tau} =2\mu \big\{
   F^{\mu\nu}S_{\nu} -u^{\mu}( u_{\nu}F^{\nu\lambda}S_{\lambda} )
   \big\} +2\epsilon \big\{ {\tilde F}^{\mu\nu}S_{\nu}
   -u^{\mu}(u_{\nu}{\tilde F}^{\nu\lambda}S_{\lambda}) \big\}.
   \end{array}\label{2} \end{equation} Here ${\tilde F}_{\mu\nu}$ is
the dual electromagnetic field tensor, and we account for possibility
of non-vanishing electric dipole moment, $\epsilon$. Note that the
second term in eq.(\ref{2}) violate $T$ invariance.
Let us generalize eq.(\ref{2}) for the case of
models with $CP$ invariance and $P$ nonconservation.
A $P$-noninvariant theory imploys
existence of a prefered direction that can be determined by the
constant vector ${\vec k}$.  The Lorentz invariant generalization of
the eq.(\ref{2}) is given by the substitution $F_{\mu\nu} \rightarrow
F_{\mu\nu} +G_{\mu\nu}$, where $G_{\mu\nu}$ is an anti symmetric
tensor organized as follows: $G_{\mu\nu}=(\xi {\vec k}, \rho {\vec
k})$ .  If one demands that the generalized equation
should be linear over field $F_{\mu\nu}$ then the only possibility
is to identify vector ${\vec k}$ with unit vector ${\vec n}$:  ${\vec
k}={\vec n}={{\vec u} \over u}$.  The quantities $\xi$ and
$\rho$ are pseudo scalar and scalar, respectively, and they do not
depend on $F_{\mu\nu}$.  In the rest frame of particle for the model
with $T$ invariance we can get
   \begin{equation}
   \begin{array}{c}
   {d{\vec S} \over d t}=2\Big[ {\vec S}\times {\vec R} \Big],~{\vec
   R}={\mu \over u_{0}}({\vec B}_{0}+\rho{\vec n})
   \end{array}\label{3}
   \end{equation}
where ${\vec B}_{0}=u_{0}{\vec B} +[{\vec E}\times{\vec u}] -{{{\vec
u}({\vec u}{\vec B})} \over {1+u_{0}}}$ is the magnetic field in the
rest frame. It follows that even in the absence of electromagnetic
field spin precession exist if the spin ${\vec S}$ is not exactly
parallel with velocity ${\vec u}$.  If for description of the
neutrino spin states we use the spin tensor $S=({\vec \sigma}{\vec
S})$ the evolution of which is determined by
$S(t)=U S(t_{0}) U^{+}$, then we can obtain~\cite{9}
   \begin{equation}
   \begin{array}{c}
   {d U \over d t}=i({\vec \sigma}{\vec R})U.
   \end{array}\label{4.1}
   \end{equation}
This equation will coincide
with one for the evolution operator that determines solutions of the
Schr\"odinger eq.(\ref{1}) if we take $\rho={u_{0} \over
\mu}({V \over 2} -{\Delta m^2A \over 4E})$ and the transversal
magnetic field
${\vec B}_{\perp}={\vec B}-{\vec n}({\vec B}{\vec n})$
is substituted by ${{\vec B}_{0} \over u_{0}}$.

   Now we consider the neutrino spin procession in a field of
electromagnetic wave with frequency $\omega$. We will denote by
${\vec e}_{3}$ the axis that is parallel with ${\vec u}$ and by
$\phi$ the angle between ${\vec e}_{3}$ and the wave vector of the
wave. The magnetic field in the rest frame is given by
   \begin{equation} \begin{array}{c}
   {\vec B_{0}}=u_{0}[(\cos{\phi}-\beta)B_{1}{\vec e}_{1}
   +(1-\beta\cos{\phi})B_{2}{\vec e}_2
   -{\sin{\phi} \over u_{0}}B_{1}{\vec e}_{3}],
   \end{array}\label{5}
   \end{equation}
where $\beta={u \over u_{0}}$, ${\vec e}_{1,2,3}$ are
the unit orthogonal vectors. For the wave of circular
polarization $B_1=B\cos{\psi}$, $B_2=B\sin{\psi}$, and the phase
$\psi=g\omega t(1-{\beta \over \beta_{0}}\cos{\phi})$ depends on
the wave speed $\beta_{0}$($\beta_{0}\leq 1$, $g=\pm 1$).Expanding
over small parameter $1/u_{0}\ll1$ and neglecting terms which are
proportional to ${\mu B \over u_{0}}$, we get
   \begin{equation}
   \begin{array}{c}
   {\vec R}=\Big({V \over 2}
   -{\Delta m^2A \over 4E}\Big){\vec e}_{3}
   -\mu B(1-\beta\cos{\phi})({\vec e}_{1}\cos{\psi}
   -{\vec e}_{2}sin{\psi})
   +O({\mu B \over u_{0}}).
   \end{array}\label{6}
   \end{equation}
In this case the solution of eq.(\ref{4.1}) can be written in the form
   \begin{equation}
   \begin{array}{c}
   U=U_{{\vec
   e}_{3}}(\psi-\psi_{0})U_{\vec l}~(\chi-\chi_{0}),~~
   U_{\vec l}~(\chi)=exp(i{({\vec \sigma}{\vec
   l})\over l}{\chi \over 2}).
   \end{array}\label{7}
   \end{equation}
The evolution operator is a combination of
the operator which
discribes rotation on angle $\chi-\chi_{0}=2l(t-t_{0})$ around the
axis ${\vec l}$, and the rotation operator on the
angle $\psi-\psi_{0}$ around the axis ${\vec e}_{3}$.  For the vector
${\vec l}$ we get
   \begin{equation}
   \begin{array}{c}
   {\vec l}=\Big({V \over 2}
   -{\Delta m^2A \over 4E}
   -{{\dot \psi} \over 2}\Big){\vec e}_{3}
   -\mu B(1-\beta\cos{\phi})({\vec e}_{1}\cos{\psi_{0}}
   -{\vec e}_{2}sin{\psi_{0}}).
   \end{array}\label{8_}
   \end{equation}
 The conversion probability among the two neutrino
states, $\nu_{L}$ and $\nu_{R}$, could became sufficient when the
vector ${\vec l}$ is orthogonal or nearly orthogonal to the axis
${\vec e}_{3}$.This will happen when the condition \begin{equation}
   \begin{array}{c}
   \Big|
   {V \over 2}
   -{\Delta m^2A \over 4E}
   -{{\dot \psi} \over 2} \Big|
   \ll\mu B(1-\beta\cos{\phi})
   \end{array}\label{8}
   \end{equation}
is valid. This unequality binds together the properties of neutrino
($\mu$, $\Delta m^2$, $E$, $\theta$) and
medium ($V$), as well as the direction of propagation and the other
characteristics of the electromagnetic wave. Condition (\ref{8})
predicts the new type of neutrino resonances $\nu_{L}\leftrightarrow
\nu_{R}$ in the electromagnetic field.

 In our previous studies \cite{2PFP,9_6}
we discussed neutrino conversion and oscillations
among the two neutrino species induced by strong twisting magnetic
field, ${\vec B}={\vec B}_{\perp}e^{i\psi (t)}$.
We introduced the critical magnetic field
${\tilde B}_{cr}(\Delta m^2,\theta,n_{eff},E, \\{\dot \psi}(t))$
as a function of the
neutrino properties and the angle $\psi$ specifying the variation of
the magnetic field in the plane transverse to the neutrino motion,
that determines the range of fields ($B\geq B_{cr}$) for which the
magnetic field induced neutrino conversion become significant. For
the critical field we got
   \begin{equation}
   \begin{array}{c}
   \tilde B_{cr}
   =\left|{1\over 2\mu}\Big({\Delta m^{2}_{\nu}A \over 2E}-
   \sqrt{2}G_{F}n_{eff} +{\dot \psi})\right|.
   \end{array}\label{Bcr}
   \end{equation}
 Remarkably, this result follows from (\ref{8}) if $\cos{\phi}=0$.
It was also pointed out~\cite{2PFP,9_6} that effects of the magnetic
field induced neutrino conversion become important if the following
two conditions are satisfied:  1)the magnetic field exceeds the
critical value $B_{cr}$ ( $B\geq B_{cr}$) and 2)the length $x$ of the
neutrinos path in the medium must be greater than the effective
oscillation length $L_{eff}$, $x\geq {L_{eff} \over 2}$.  We used this
criterion to get constraints on $\mu$. In
particular, to avoid the loss of a substantial amount (25$\%$) of
energy that could escape during the supernova explosion~\cite{F} together
with the sterile neutrinos $\nu_{eR}$, we has to constrain the
magnetic moment on the level of $\mu\leq 10^{-11}\mu_{0}$. More stringent
constraint $\mu\leq 10^{-15}\mu_{0}$ is also obtained~\cite{10}
from consideration of the magnetized neutron
star cooling.

One of the authors (A.S.) should like to thank J.Dias de Deus
and A.Mourao for hospitality during the Meeting
on New Worlds in Astroparticle Physics and
J.Pulido for fruitful discussions.

\end{document}